# Effect of substrate topography, material wettability and dielectric thickness on reversible electrowetting[1]

**Nikolaos T. Chamakos**[a], **George Karapetsas**[b] and **Athanasios G. Papathanasiou**[a]*

a) *School of Chemical Engineering, National Technical University of Athens, 15780, Greece*
b) *Department of Chemical Engineering, Aristotle University of Thessaloniki, 54124, Greece*
*E-mail: pathan@chemeng.ntua.gr

**Abstract**

Recent experiments by Kavousanakis et al., *Langmuir*, 2018 [1], showed that reversible electrowetting on superhydrophobic surfaces can be achieved by using a thick solid dielectric layer (e.g. tens of micrometers). It has also been shown, through equilibrium (static) computations, that when the dielectric layer is thick enough the electrostatic pressure is smoothly distributed along the droplet surface, thus the irreversible Cassie to Wenzel wetting transitions can be prevented. In the present work we perform more realistic, dynamic simulations of the electrostatically-induced spreading on superhydrophobic surfaces. To this end, we employ an efficient numerical scheme which enables us to fully take into account the topography of the solid substrate. We investigate in detail the role of the various characteristics of the substrate (i.e. the dielectric thickness, geometry and material wettability) and present relevant flow maps for the resulting wetting states. Through our dynamic simulations, we identify the conditions under which it is possible to achieve reversible electrowetting. We have found that not only the collapse (Cassie-Baxter to Wenzel) transitions but also the contact angle hysteresis of the substrate significantly affects the reversibility.

**Introduction**

The dynamic control of the apparent wettability of superhydrophobic surfaces has lately attracted strong scientific interest [2, 3] since it is related with promising technological applications which may involve liquid motion without moving mechanical parts (e.g. in lab-on-a-chip devices) [4]. Wettability modification, on geometrically structured surfaces, is, however, commonly accompanied by wetting transitions, i.e. from a Cassie-Baxter state, where the liquid is suspended above the solid protrusions, to a Wenzel-type state where the liquid, penetrates the solid roughness [4]. Then the mobility of the droplet is considerably limited.

---

[1] Published in Colloids and Surfaces, A: Physicochemical and Engineering Aspects (https://doi.org/10.1016/j.colsurfa.2018.07.043). This article is licensed under a CC-BY-NC-ND license.

Modification of the apparent wettability of a solid surface can be realized by a plethora of techniques including pH [5, 6] and temperature variation [7-9], light illumination on photo-responsive surfaces [10-13] as well as transitions occurring by surface morphology modification [14]. The above are typically termed as ex-situ wettability switching techniques where a different liquid droplet is required to study the wettability response before and after the surface treatment. Such ex-situ methods, however, are inappropriate for miniaturized devices e.g. a medical lab-on-a-chip where a single droplet of blood must be transferred through a series of sensors and micro-reactors.

The above applications require the so called in-situ techniques where the Cassie-Baxter and the Wenzel states can be reversibly switched. An example of an in-situ wettability switching technique, at an oil-water-solid system, is the redox reaction of conducting polymer films [15]. In particular, the liquid-solid adhesion can be controlled by oxidizing and reducing a polypyrrole (PPy) substrate. A more versatile technique, however, which can be used on common water-air-solid systems, is electrowetting (see refs. [16, 17] and references therein). In the latter method, the solid wettability is electrostatically enhanced by applying a voltage between a base electrode, which is coated by a dielectric layer, and droplet of a conductive liquid. Despite that electrowetting-induced Cassie-Baxter to Wenzel transitions are easily realizable, the reverse are extremely challenging to be performed spontaneously [18]; the reverse transition may require rapid heating of the solid substrate [19]. A plethora of studies lately have focused on realizing reversible Cassie-Baxter to Wenzel transitions [20-22], with, however, disputable results regarding: the suitability of the solid topography for technological applications [23], or the wetting reversibility range [24].

Recently, Kavousanakis et al. [1], through theoretical computations supported by experiments, showed that fully reversible electrowetting can be realized when the dielectric thickness is sufficiently large, provided that the surface texture is such that it exhibits high resistance in impalement transitions. In contrast, when the dielectric is made thin, the same surface structures (fabricated by means of colloidal lithography and plasma etching) [25] could not perform reversible electrowetting even when a particular surface type with high resistance to impalement transitions has been used. Nevertheless, the critical dielectric thickness above which reversibility is observed can be affected significantly by the specific topography of the solid substrate. This study however has raised some interesting questions: a) What is the dynamics of the wetting transition during the electrowetting phenomenon? b) Based on the dynamics, which type of topography structures actually promote reversible electrowetting and most importantly, c) Is it possible to provide a theoretical prediction of the critical dielectric thickness for fully reversible electrowetting?

The aim of the current work is to shed light on the mechanism of wetting reversibility by performing realistic time-dependent simulations of electrowetting, taking into account important characteristics of the solid substrate, i.e. the surface geometry, the material wettability and the dielectric thickness. To this end, we employ an efficient sharp-interface, continuum-level formulation for modelling the motion of liquid droplets on structured surfaces, where the liquid-vapor and the liquid-solid interfaces of the droplet are treated in a unified context (one equation for both interfaces) [26, 27]. This is achieved by using a disjoining pressure term, modeling the liquid-solid micro-scale interactions, and thus avoiding the implementation of any boundary condition at the contact line(s). The model that has been developed allows, without making any simplification concerning the droplet shape or the field distribution, to investigate the electrostatically-induced spreading on superhydrophobic surfaces and the accompanied transitions between different wetting states. The present article is organized as follows: we first present the mathematical framework used for our simulations. Next, our numerical results regarding electrowetting-induced wetting transitions are presented and discussed. Concluding annotations are made in the final section.

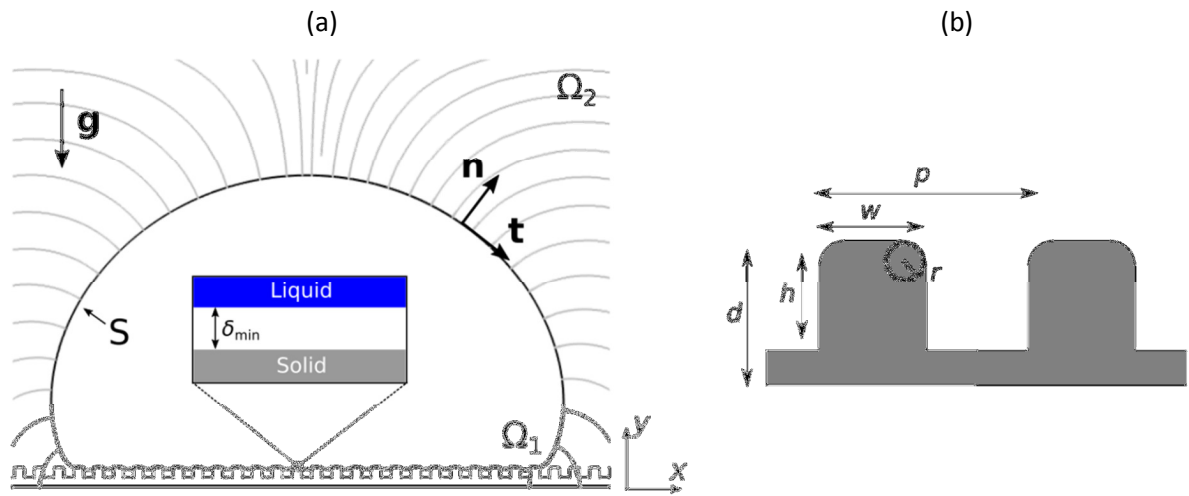

**Figure 1:** (a) Schematic of the electrowetting setup of a droplet on a structured dielectric substrate. (b) Geometric parameters of the structured solid surface.

**Problem formulation**

We consider the dynamics of a 2D droplet of a conductive liquid deposited on a dielectric layer coating a structured electrode when subjected to an electric voltage, i.e. a typical electrowetting-on-dielectric (EWOD) setup (see Figure 1a). The fluid is considered to be an incompressible Newtonian liquid with constant density, $\rho$, and viscosity, $\mu$. The geometric characteristics of the dielectric, which we consider in this work, are presented in Figure 1b. The dynamics of the liquid droplet are governed by the Navier-Stokes equations, i.e. the conservation of mass and momentum, given below:

$$\rho\left(\frac{d\mathbf{u}}{dt}+\mathbf{u}\cdot\nabla\mathbf{u}\right)=-\nabla p^L+\mu\nabla^2\mathbf{u}+\rho\mathbf{g}, \tag{1.1}$$

$$\nabla\cdot\mathbf{u}=0,$$

where, $\mathbf{u}=(u_x,u_z)$ and $p^L$ are the fluid velocity field and pressure, respectively, and $\mathbf{g}$, denotes the gravitational acceleration.

For the scope of the current work, we employ a model which has proven to be very efficient for the study of droplet's static and dynamic behavior on structured solid surfaces [26, 27]. According to this scheme, the liquid-vapor and the liquid-solid interfaces of the droplet are treated in a unified context (one equation for both interfaces). Therefore, the solution of the Navier-Stokes equations (Eq. 1.1) is determined subject to a single stress balance boundary condition applied at the whole droplet surface (S in Figure 1a), referred from now on as the liquid-ambient interface. In particular, as described in detail in our previous works (see [26] and [27]), the liquid-solid interactions are lumped in a disjoining pressure term, $p^{LS}$, which will now be included in the normal component of the interface force balance:

$$\tau_{nn}|_{liquid}=\Delta p-\gamma_{LA}C-p^{LS}-p_{el}, \tag{1.2}$$

where $C$ is the local mean curvature, $\Delta p$ is the pressure jump across the interface, $\gamma_{LA}$ is the liquid-ambient interfacial tension, $p_{el}$ is the electrostatic pressure due to the effect of the electric field, and $\tau_{nn}$ is the normal stress. In the above equation, $\tau_{nn}=\mathbf{n}\cdot\boldsymbol{\tau}\cdot\mathbf{n}$ where $\boldsymbol{\tau}$ is the viscous stress tensor ($\boldsymbol{\tau}=\mu\left[\nabla\mathbf{u}+(\nabla\mathbf{u})^T\right]$) and, $\mathbf{n}$, the unit normal of the liquid-ambient interface (see Figure 1a). The disjoining pressure, $p^{LS}$, is defined as the pressure in excess of the external pressure that must be applied to a fluid between two plates to maintain a given separation distance, that is essentially, the force of attraction or repulsion between the plates per unit area [28]. We formulate the disjoining pressure according to the following expression [29]:

$$\frac{R_o}{\gamma_{LA}}p^{LS}=w^{LS}\left[\left(\frac{\sigma}{\delta/R_o+\varepsilon}\right)^{C_1}-\left(\frac{\sigma}{\delta/R_o+\varepsilon}\right)^{C_2}\right], \tag{1.3}$$

which resembles a Lennard-Jones type potential. Alternative formulations for the disjoining pressure could also be employed, as demonstrated in[30]. In the above equation, the depth of the potential well is proportional to a wetting parameter, $w^{LS}$, which is directly related with the solid wettability (an increase of $w^{LS}$ results in a deeper well of the potential, indicating stronger liquid-solid affinity). In addition, the exponents $C_1$ and $C_2$ control the range of the molecular interactions (large $C_1$ and $C_2$ reduce the range within which these interactions are active. The distance, $\delta$, between the liquid and the solid surface determines whether the disjoining pressure is attractive (modeling van der Waals interactions, for relatively large $\delta$) or repulsive (modeling steric forces and electrostatic interactions determined by an overlapping of the electrical double layers, for small $\delta$) [28]. In the case of a perfectly flat solid surface, the distance, $\delta$, is defined as the vertical distance of the liquid surface from the solid boundary. For non-flat, rough, solid surfaces, the definition of distance, $\delta$, requires special consideration. Here, we take, $\delta$, as the Euclidean distance from the solid. This quantity is obtained by solving the Eikonal equation[31], which expresses the signed distance from a boundary (even arbitrarily shaped). In our formulation we consider that the liquid and the solid phases are separated by an intermediate layer (with thickness $\delta_{min}$) which is stabilized by the presence of the disjoining pressure (see Figure 1a). In particular at $\delta = \delta_{min}$ the repulsive and attractive forces balance each other; further reduction of the intermediate layer thickness, below $\delta_{min}$, would generate strong repulsion. The minimum allowed liquid-solid distance $\delta_{min}$ is determined by the constants $\sigma$ and $\varepsilon$. Specifically, for $\delta = \delta_{min} \Leftrightarrow p^{LS} = 0 \Leftrightarrow \delta_{min} = R_o(\sigma - \varepsilon)$.

Regarding the tangential stress component along the liquid surface, we will use a Navier slip model with an effective slip coefficient, $\beta_{eff}$, active only in close proximity to the solid:

$$\tau_{nt} \big|_{liquid} = \beta_{eff} (\mathbf{t} \cdot \mathbf{u}), \tag{1.4}$$

where $\tau_{nt} = \mathbf{n} \cdot \boldsymbol{\tau} \cdot \mathbf{t}$ denotes the shear stress; $\mathbf{t}$ denotes the unit tangent of the liquid-ambient interface (see Figure 1a). In the above, a uniform interfacial tension along the interface has been considered ($\nabla_s \gamma_{LA} = 0$). The Navier slip model is active only in the vicinity of the solid surface, and this is achieved by using an effective slip coefficient, $\beta_{eff}$, of the following form:

$$\beta_{eff} = \frac{\mu \beta_{LS}}{R_0} \left( 1 - \tanh\left[ p_{trs} \left( \frac{\delta}{\delta_{min}} - 1 \right) \right] \right). \tag{1.5}$$

Here, the dimensionless slip parameter, $\beta_{LS}$ (i.e. scaled inverse slip length), regulates the shear strength of the liquid on the solid surface. The above formulation is a simple way to denote in a continuous manner the transition from a shear-free boundary condition, applied on the liquid-ambient interface, to a partial slip boundary condition along the liquid-solid interface. In particular, in the limit $\delta \approx \delta_{min}$, the above equation reduces to $\beta_{eff} = \mu \beta_{LS} / R_0$, whereas for $\delta > \delta_{min}$, yields $\beta_{eff} = 0$ and thus the tangential stress balance reduces to a shear-free boundary condition. The parameter, $p_{trs}$, ensures a sharp transition between these two regimes. We note that, in the computations presented in this paper, we assume $p_{trs}$ = 5. Finally, we consider that typical values of the dimensionless slip parameter, $\beta_{LS}$, are of the order of $R_0 / \delta_{min}$.

The effect of the electric field is incorporated in the normal interfacial stress balance through the electrostatic pressure term, $p_{el}$, which acts on the liquid surface, with a negative contribution to the total pressure [16]. The electrostatic pressure is given by $p_{el}$ = $\varepsilon_0 E^2/2$, where $\varepsilon_0$ is the vacuum permittivity ($\varepsilon_0$ = 8.854 × 10$^{-12}$ F/m) and $E$ the electric field strength. The electric field strength, $E$, is calculated along the droplet surface by solving the equations of electrostatics (Gauss' law for electricity) for both the ambient phase and the dielectric material:

$$\nabla \cdot ( \varepsilon_r \nabla \varphi) = 0, \tag{1.6}$$

Where, $\varphi$, is the electric potential. Eq. (1.6) is not solved inside the droplet since the droplet is considered conductive. For simplicity, the permittivity, $\varepsilon_r$, is assumed to be given by a continuous function of this form, $\varepsilon_r = (\varepsilon_s - \varepsilon_d) \tanh(a_{trs} \delta) + \varepsilon_d$. According to this expression, the permittivity, $\varepsilon_r$ becomes equal to, $\varepsilon_s$, in the ambient phase (insulating medium) and equal to, $\varepsilon_d$, for the solid dielectric, respectively. When $a_{trs}$ acquires a high value, a sharp transition between the two regions is achieved. For the simulations that will be presented below, $a_{trs} = 500$ is assumed. Equation (1.6) is solved accounting for the following boundary condition at the liquid-ambient interface (S in Figure 1a):

$$\varphi = V, \tag{1.7}$$

where, $V$, is the voltage applied between the base electrode and the conductive droplet. Moreover, at the bottom of the solid dielectric (base electrode) we apply:

$$\varphi = 0. \tag{1.8}$$

As measure of the strength of the electric field we consider the dimensionless electrowetting number, $\eta = \frac{\varepsilon_0 \varepsilon_r V^2}{2d\gamma_{LA}}$, which expresses the relative strength of the electrostatic over the surface tension forces in the system, assuming a uniform electric field at the liquid-solid interface (ideal parallel plate capacitor). Finally, the following kinematic boundary condition is imposed along the liquid-ambient interface:

$$(\mathbf{u}_{mesh} - \mathbf{u}) \cdot \mathbf{n} = 0, \tag{1.9}$$

where $\mathbf{u}_{mesh}$ is the velocity of the mesh at the interface. The above model has been implemented in COMSOL Multiphysics® commercial software.

**Results**

In this work we model the electrowetting dynamics of a 2D droplet on a multi-striped dielectric which is presented in Figure 1b. For the purposes of our study we will examine the flow dynamics of a glycerin/water mixture droplet (85% of glycerin with $\rho$ = 1275 kg/m³, $R_0$ = 1.5 mm, $\gamma_{LA}$ = 0.07 N/m and $\mu$ = 116 mPa s) resting on a geometrically structured solid dielectric; examples of typical solid structures that we have considered are presented in Figure 2. In Fig 2a a solid with stripes having width $w$ = 75 um and pitch $p$ = 150 um is presented. In Fig. 2b and 2c we consider structures with either reduced width ($w$ = 30 um) and same pitch ($p$ = 150 um) or reduced pitch ($p$ = 105 um) but same width ($w$ = 75 um), respectively. In all simulations that will be presented below the relative permittivity $\varepsilon_d$ of the solid dielectric is 3.8, while $\varepsilon_s$ is considered to be equal to 1.

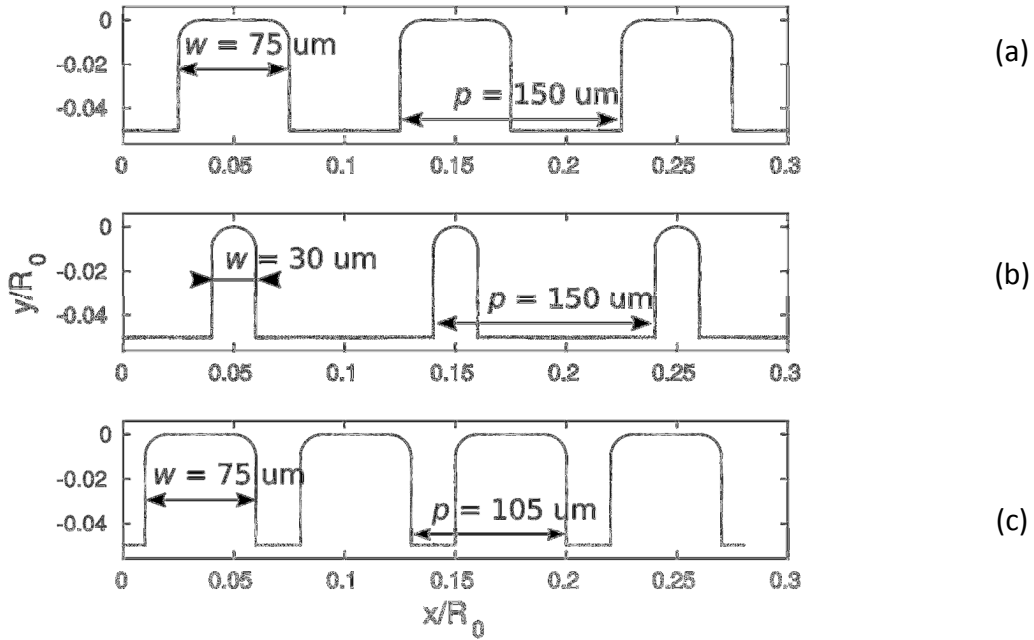

**Figure 2:** Different solid structures used in the electrowetting simulations. The corresponding geometric parameters are the following: (a) $h$ = 75 um, $w$ = 75 um, $p$ = 150 um, $r$ = 15 um, (b) $h$ = 75 um, $w$ = 30 um, $p$ = 150 um, $r$ = 15 um and (c) $h$ = 75 um, $w$ = 75 um, $p$ = 105 um, $r$ = 15 um.

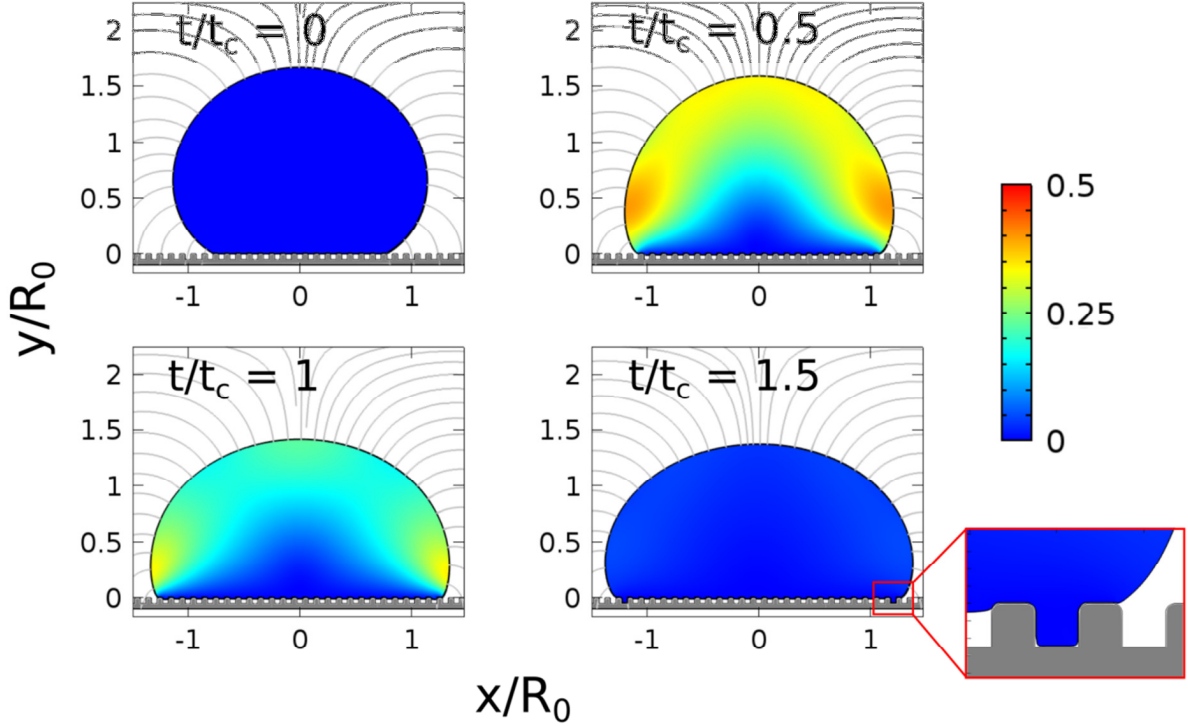

**Figure 3:** Visualization of the normalized velocity magnitude (from $t/t_c$ = 0 to $t/t_c$ = 1.5, where the liquid has effectively come to rest) of a glycerin/water droplet on a structured solid dielectric ($\theta_Y$ = 120°, $\varepsilon_d$ = 3.8, $h$ = 75 um, $w$ = 75 um, $p$ = 150 um, $r$ = 15 um and $d$ = 150 um). A voltage of 792 V ($\eta$ = 1) is applied at $t/t_c$ = 0. The electric field lines are also depicted. As observed in the inset, only the outer grooves are filled with liquid, however, a Cassie-like state is observed elsewhere. The disjoining pressure parameters we use are, according our previous work (see [26] and [27]): $C_1$ = 12, $C_2$ = 10, $\sigma$ = 9×10$^{-3}$ and $\varepsilon$ = 8×10$^{-3}$ (resulting in a $\delta_{min}$ = 1.5 um) while the dimensionless slip parameter: $\beta_{LS}$ = 10$^3$.

*Electrowetting dynamics on structured solid dielectrics*

In our initial frame, we assume that the droplet rests at equilibrium on the structured solid surface; we find this initial state of equilibrium by letting a spherical droplet spread along the solid surface. At $t$ = 0$^+$, a voltage, $V$, is applied between the droplet and the base electrode, triggering an electrostatically-induced spreading; the dimensionless electrowetting number is set to $\eta = 1$. Indicative snapshots of the droplet profiles at selected time instances, as the liquid spreads out on the solid topography presented in Figure 2a, are demonstrated in Figure 3. In this figure we visualize the normalized velocity magnitude from $t/t_c$ = 0 to $t/t_c$ = 1.5, where the liquid has effectively come to rest; the characteristic time equals to $t_c = \sqrt{\dfrac{R_0}{g}}$ = 0.0124 s. The dielectric thickness (from the apex of the solid protrusions to the dielectric base) is $d$ = 150 um (see also Figure 1). In the current simulation we consider a solid material with Young contact angle equal to 120°. We observe that in the absence of an electric field and for the specific geometric characteristics and solid wettability, the droplet equilibrates in a Cassie-Baxter state (air pockets are trapped beneath the droplet). When the voltage is applied the droplet spreads out to find its new equilibrium state. At the early stages, the droplet does not wet the asperities of the solid remaining in Cassie-Baxter state. However, at some point (at $t/t_c$=1.5) the interfacial tension is no longer able to sustain the local electrostatic pressure ($\varepsilon_0 E^2/2$) and an impalement transition takes place at the outer grooves of the solid surface covered by the droplet (see the inset of Figure 3 as well as the corresponding video clip included in the supplementary material). Past these transitions the droplet effectively comes to rest.

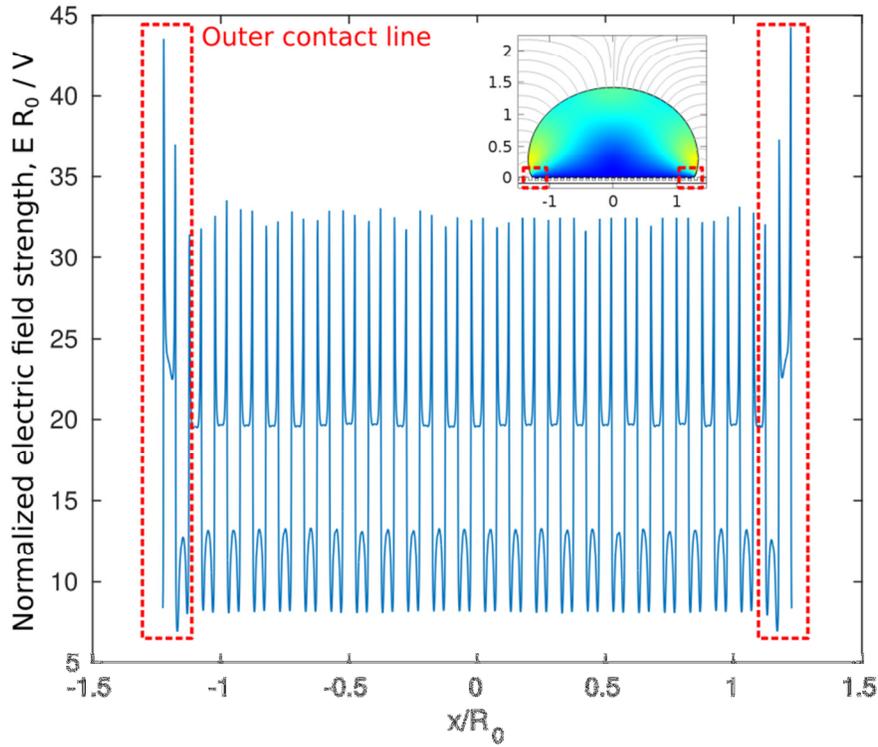

**Figure 4:** Variation of the normalized electric field strength, $\frac{E R_0}{V}$, along the effective liquid-solid interface of a glycerin/water droplet on the structured solid dielectric presented in Figure 2a for $\eta$ = 1 and $t/t_c$ = 1 ($\theta_Y$ = 120°, $\varepsilon_d$ = 3.8, $d$ = 150 um). The normalized electric field strength reaches its maximum value at the outer region of the liquid-solid interface (see also the inset of the Figure).

The local switch from a Cassie-Baxter to a Wenzel state, observed only in the vicinity of the outer contact line of the droplet, indicates that the electrostatic pressure, and thus the electric field value, at this region should be maximal. This can be actually illustrated by plotting the normalized electric field strength, $\frac{E\,R_0}{V}$, distribution along the effective liquid-solid interface at $t\,/\,t_c$ = 1 (see Figure 4). In particular, we observe that, $\frac{E\,R_0}{V}$, reaches a maximal value at the outer contact line region from where the collapse transition initiates (see also the inset of Figure 4). Such a phenomenon, where Wenzel-like states are observed at the outer region of the effective liquid-solid interface, whereas Cassie-like states are detected at the inner region, has been also reported in the experimental work of Manukyan et al. [32].

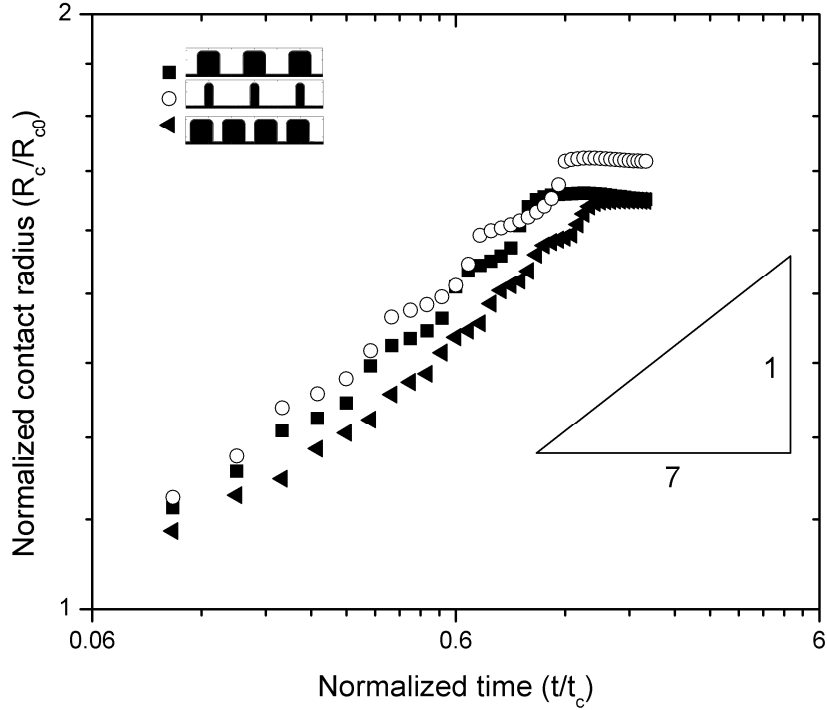

**Figure 5:** Temporal evolution of the normalized contact radius of a droplet spreading, due to electric field application, on the structured solid surfaces presented in Figure 2. Snapshots of the droplet, on the solid surface presented in Figure 2b, are demonstrated in Figure 3.

In order to investigate the effect of solid geometry on electrospreading we have also performed simulations for the other two topographies presented in Figure 2 (2b and 2c where the stripes width and distance (pitch) has been reduced, respectively) while keeping the same dielectric thickness value ($d$ = 150 um), wettability of the material ($\theta_Y = 120^o$) and electrowetting number ($\eta = 1$). In Figure 5 we present the temporal evolution normalized contact radius ($R_c/R_{c0}$, where $R_{c0}$ is the contact radius at $t$ = 0) of the droplet for all the solid structure cases. We observe that contact radius grows with time according to the power law of ~1/7 which is consistent with the predictions of the Tanner law [33] in the case of a 2D droplet. Such a behavior has been also experimentally observed for fluids spreading on smooth as well as structured substrates, with or without electrostatic assistance [7, 34, 35]. As it is clearly shown in this figure, the droplet essentially reaches its maximum extent of spreading at $t/t_c \sim 1$.

Despite that the contact radius grows with time according to the same power law for all the solid structures, it would be interesting here to study the details of the resulting droplet profile at equilibrium for each case since it could greatly affect the eventual droplet mobility (e.g. a Cassie-Baxter and a Wenzel wetting state may exhibit the same apparent contact angle but a significantly different liquid-solid friction coefficient).

The initial as well as the equilibrium droplet profiles (at $t/t_c$ = 10) after applying electric field, for the substrate topographies presented in Figure 2b and c, along with contour lines of the electric field are shown in Figure 6. We observe that although in all cases the droplet was initially in a Cassie-Baxter state, the final equilibrium wetting state depends on the solid topography characteristics. In particular, a variety of equilibrium states, including a Wenzel (Figure 6a2), a Cassie-Baxter (Figure 6b2), and a mixed wetting state (as discussed in the previous paragraph, see also Figure 3), are obtained. In the fully collapsed case, for the topography presented in Figure 2b, the transition sets is simultaneously (at $\sim t/t_c$ = 0.1) for all the wetted grooves (see also the corresponding video clip included in the supplementary material); this transition is similar to the pressure-driven collapse of conventional superhydrophobic surfaces. The latter is attributed to the low stability limit of the Cassie-Baxter state (the locally developed electrostatic pressure cannot be sustained by the interfacial tension) as a result of the sparsely spaced grooves in this case.

The final wetting state of the droplet depends on the interplay between the electrostatic pressure and capillary forces. On the one hand, the capillary forces are regulated by the geometric characteristics of the substrate topography. Thus as previously discussed, for a given dielectric thickness and strength of the electric field, the Wenzel state (see Figure 6b2) is favored by decreasing the width of the stripes (for small $w$ value as in the topography presented in Figure 2b) whereas the Cassie-Baxter state (see Figure 6a2) is favored by decreasing the distance between the stripes (for small $p$ value as in the topography presented in Figure 2c). Intermediate values of stripes width and pitch may result in mixed wetting states, where the liquid has partially penetrated the solid roughness, as presented in Figure 3.

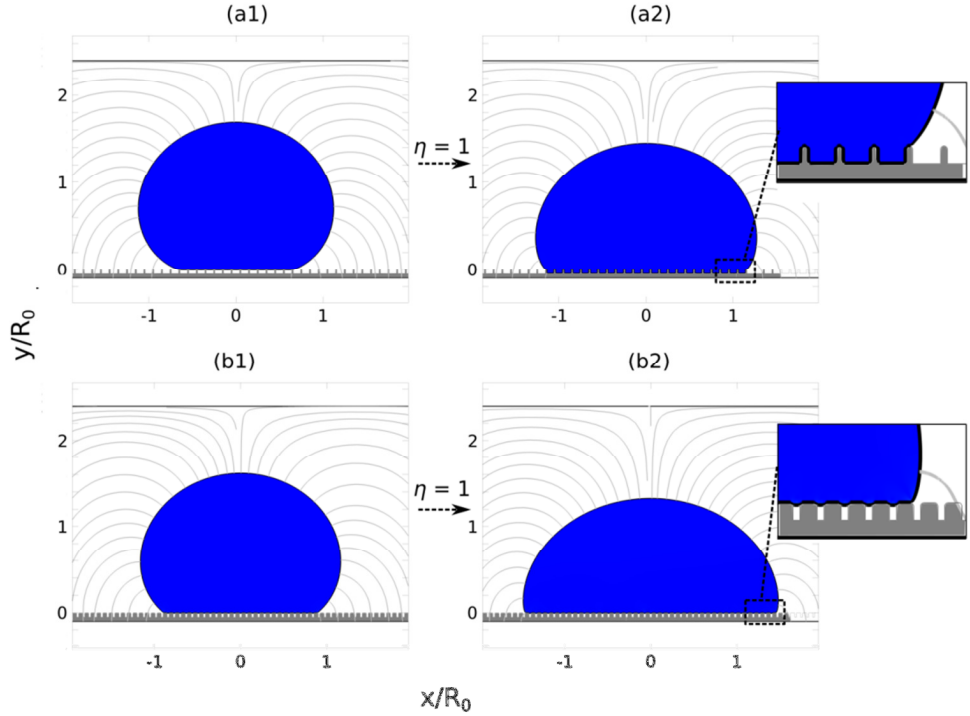

**Figure 6:** Initial (a) and final (b) wetting states (at $t/t_c$ = 10) of a droplet on the various solid structures demonstrated at Figure 2b and c ($\theta_Y$ = 120°, $\varepsilon_d$ = 3.8, $\eta$ = 1 and $d$ = 150 um for all cases). (a2) a Wenzel (at the solid structure presented in Figure 2b) as well as (b2) a Cassie-Baxter (at the solid structure presented in Figure 2c), is observed, as a result of the different geometric characteristics. Video clips of the droplet dynamic behavior on all the various solid structure cases are included in the supplementary material.

The electric field on the other hand, can be significantly affected by both the thickness of the dielectric layer and the electrowetting number, $\eta$, which is measure of the strength of the electric field. To quantify these effects we next present a parametric analysis of the equilibrium wetting state at $t/t_c$ = 10 (either Cassie-Baxter, Wenzel or mixed) with regard to these two factors (substrate geometry and electric field).

*Effect of solid topography, dielectric thickness and material wettability on collapse transitions*

In Figure 7 we present the final contact radius of the droplet, $R_c$, (considering that equilibrium has been reached at $t/t_c$ = 10), normalized by the initial contact radius, $R_{c0}$, (at $t$ = 0), for various electrowetting numbers, $\eta$, and three different dielectric thicknesses, $d$ = 90, 150 and 300 um (i.e. a thinner and thicker one than the previously discussed example). In the same figure, we also plot the effective arc-length of the liquid-solid interface at equilibrium ($t/t_c$ =

10), $A_{ls}$, normalized by the arc-length of the initial liquid-solid interface, $A_{ls0}$ (at $t = 0$). This ratio can be considered as a measure of the coverage of the solid asperities by the liquid or the liquid-solid contact; high values correspond to Wenzel state whereas for values close to 1 the system exhibits a Cassie-like state. The geometric parameters in this case correspond to: $h$ = 75 um, $w$ = 75 um, $p$ = 150 um and $r$ = 15 um (see also Figure 1a), while the Young contact angle is, $\theta_Y = 120^o$.

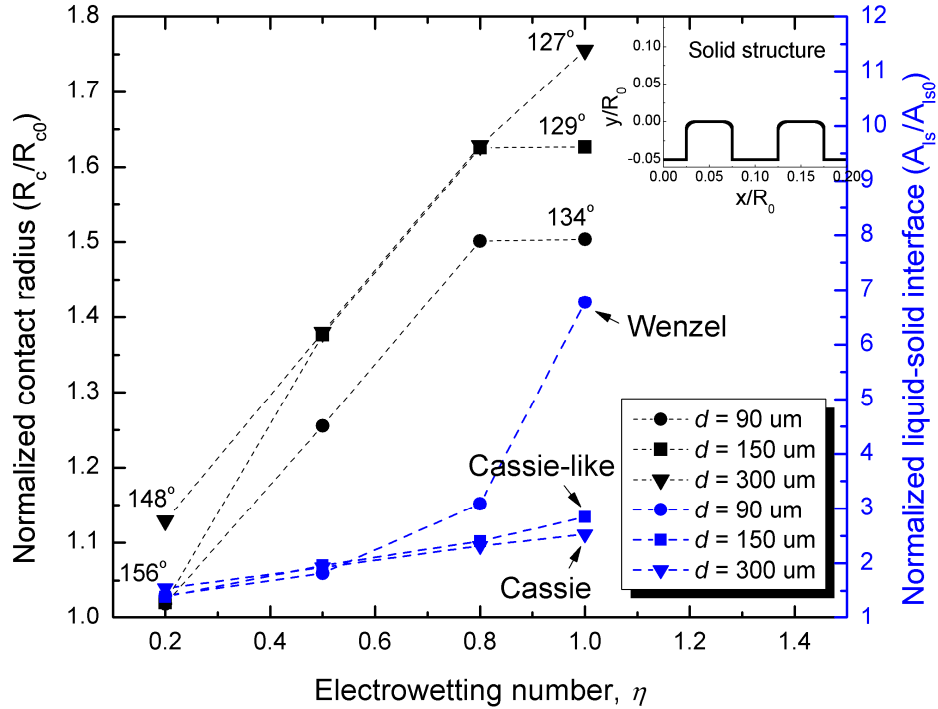

**Figure 7:** Normalized contact radius (black points) and liquid-solid interface (normalized the arc-length) (blue points) of a droplet equilibrating (at $t/t_c$ = 10) on a structured substrate ($\theta_Y = 120^o$, $h$ = 75 um, $w$ = 75 um, $p$ = 150 um, $r$ = 15 um and $d$ = 90, 150, 300 um), at various electrowetting numbers (ranging from $\eta$ = 0.2 to $\eta$ = 1). This solid surface is also presented in Figure 2a. The corresponding apparent contact angles values (for $\eta$ = 0.2 and $\eta$ = 1), at equilibrium, are also depicted on the figure. The disjoining pressure parameters we use are: $C_1$ = 12, $C_2$ = 10, $\sigma$ = 9×10$^{-3}$ and $\varepsilon$ = 8×10$^{-3}$.

Considering that according to the Lippmann equation [16, 17] a specific electrowetting number results in the same apparent contact angle for any thickness of the dielectric layer (with higher capacitance and low voltage at the thinner case, and lower capacitance and high voltage at the thicker case, respectively), similar results would be expected for the two dielectric thickness cases, at the same $\eta$. In Figure 7, however, we observe that the equilibrium deformation depends on the thickness of the dielectric. The final spreading radius for the thick dielectric layer (e.g. for $d$ = 300 um) is larger than the one of the thin layer ($d$ = 90 um). This effect is particularly noticeable for high electrowetting numbers, in our case when $\eta > 0.8$. Specifically, we observe an almost linear dependence of the normalized contact radius on electrowetting number for $\eta > 0.8$ with, however, a different slip according to the dielectric thickness. In addition, the normalized liquid-solid interface, $A_{ls}/A_{ls0}$, increases sharply for the thinner dielectric case at $\eta$ = 1, indicating that a collapse (Cassie-Baxter to Wenzel) transition occurred and the grooves of the solid substrate have been filled with liquid. On the contrary, such a transition is not observed for the thicker dielectric case. The above argument is in line with our previous theoretical [36] as well as experimental work [1, 18] claiming that the collapse transition can be avoided and thus the contact angle reversibility is feasible above a critical solid substrate thickness.

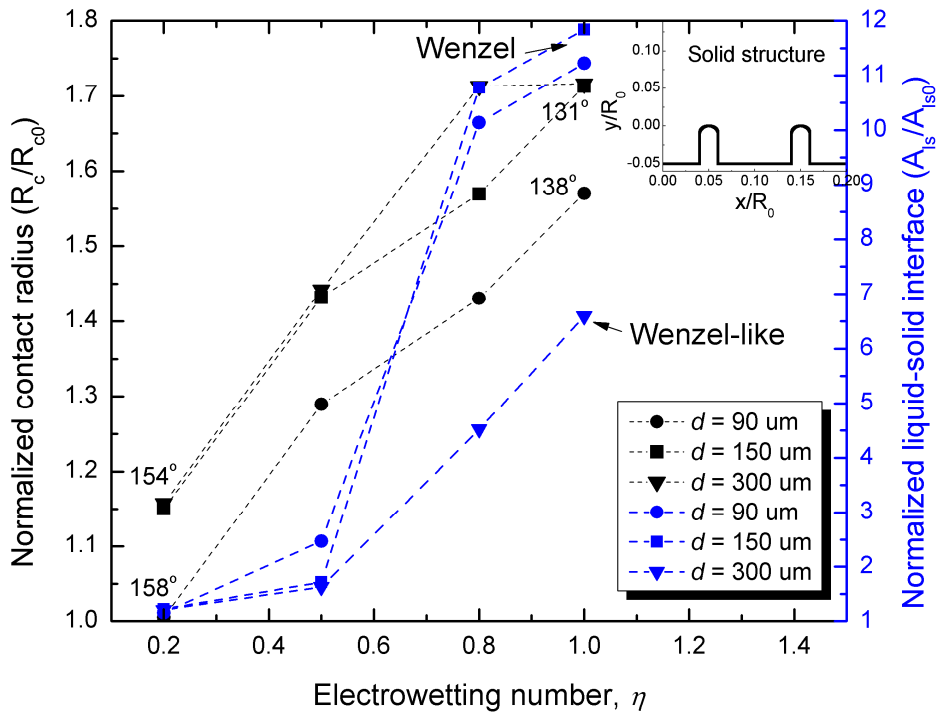

**Figure 8:** Normalized contact radius (black points) and liquid-solid interface (blue points) of a droplet equilibrating (at $t/t_c$ = 10) on a structured substrate with decreased asperities width ($\theta_Y = 120^o, h$ = 75 um, $w$ = 30 um, $p$ = 150 um, $r$ = 15 um and $d$ = 90, 150, 300 um), at various electrowetting numbers (ranging from $\eta$ = 0.2 to $\eta$ = 1). This solid surface is also presented in Figure 2b. The corresponding apparent contact angles values (for $\eta$ = 0.2 and $\eta$ = 1), at equilibrium, are also depicted on the figure.

In Figure 8 we present the normalized contact radius of a droplet equilibrating on a structured substrate with decreased pillar width, $w$, compared to that of Figure 7; the width of the protrusions here is $w$ = 30 um (see Figure 2b) while the period of the asperities (pitch), $p$, is kept constant. The normalized liquid-solid interfacial length is also plotted in the same figure. Contrary to the previous case, where the width of the protrusions is relatively larger, here a Wenzel (or Wenzel-like) state is obtained for a wide range of dielectric thicknesses. In particular, due to the decreased width of the grooves, which is given by $p - w$, the effect of capillary pressure becomes less significant and cannot support the interface against the electrostatic pressure that it experiences and resulting to the impalement transition, and thus in a Wenzel state with decreased droplet mobility. An investigation of the effect of the solid geometry on the collapse (Cassie to Wenzel) transition feasibility has also been performed in the case where the asperities distance is decreased ($p$ = 105 um as shown in Figure 9). Interestingly, we observe that no collapse transition occurs in this case, whereas the droplet remains suspended on top of the solid stripes for all the dielectric thickness cases.

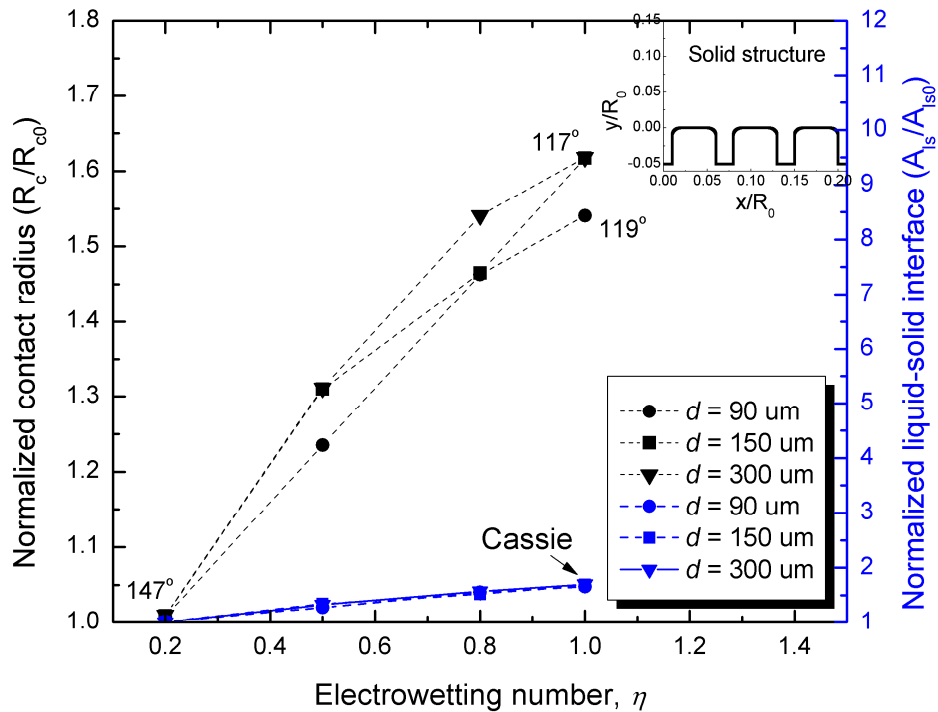

**Figure 9:** Normalized contact radius (black points) and liquid-solid interface (blue points) of a droplet equilibrating (at $t/t_c$ = 10) on a structured substrate with decreased asperities distance ($\theta_Y = 120^o$, $h$ = 75 um, $w$ = 75 um, $p$ = 105 um, $r$ = 15 um and $d$ = 90, 150, 300 um), at various electrowetting numbers (ranging from $\eta$ = 0.2 to $\eta$ = 1). This solid surface is also presented in Figure 2c. The corresponding apparent contact angles values (for $\eta$ = 0.2 and $\eta$ = 1), at equilibrium, are also depicted on the figure.

A more complete picture of the possible wetting states that can be encountered is given in Figure 10. In particular we demonstrate the existence of plethora of wetting states by plotting the contour lines of the fraction of grooves filled with liquid (number of filled grooves normalized by the total number of grooves underneath the droplet at equilibrium ($t/t_c$ = 10)) over a wide range of solid structure cases. The isoline where the above fraction equals to 1 indicates a fully collapsed state (Wenzel) whereas a value equals to zero represents a fully suspended state (Cassie-Baxter). Intermediate values correspond to mixed wetting states. Thus, an investigation is performed by modifying the following control parameters: the electrowetting number, $\eta$, the dielectric thickness, $d$, the stripes width, $w$, the pitch, $p$, and the Young contact angle, $\theta_Y$. Specifically, the fraction of grooves filled with liquid is presented in Figure 10a as a function of the dielectric thickness (ranging from $d$ = 90 um to $d$ = 150 um) and the electrowetting number (ranging from $\eta$ = 0.5 to $\eta$ = 1) for a substrate with $w$ = 75 um and $p$ = 150 um. Figure 10b presents the dependence of the dielectric thickness of stripes of different widths (ranging from $w$ = 90 um to $w$ = 150 um) for $\eta$ = 1 and $p$ = 150 um, while the effect of the period of the solid structures (ranging from $p$ = 105 um to $p$ = 150 um) for $\eta$ = 1 and $w$ = 75 um is examined in Figure 10c; note that $\theta_Y$ = 120° in the cases presented in Figures 9a, 9b and 9c. The above results show a clear connection between the solid geometry and the critical dielectric thickness beyond which no collapse transition is observed. The existence of a large number of mixed wetting states, which cannot be characterized as ideal Wenzel or Cassie-Baxter states, is in line with our previous work [36] which is based on static electrowetting computations. From these detailed flow maps, we may deduce that in order to achieve a wetting state with increased droplet mobility, i.e. an ideal Cassie-Baxter state, one has to either use a dielectric layer with large thickness or a solid substrate with very dense structures.

The intrinsic wettability of the solid surface (Young's contact angle) has also taken into consideration. To examine the effect of this parameter we produced a map (Figure 10d) by varying the Young contact angle (ranging from $\theta_Y$ = 110° to $\theta_Y$ = 140°) for $\eta$ = 1, $w$ = 75 um and $p$ = 150 um as well as the solid dielectric thickness (ranging from $d$ = 90 um to $d$ = 150 um). Although it is known that Young's contact angle cannot exceed the 120° on flat and smooth solid substrates (the wettability of PTFE), larger apparent Young contact angles can be commonly observed on dual-scale structured superhydrophobic [37]. Thus, the study of cases with $\theta_Y$ > 120° is of practical interest in the case of superhydrophobic surfaces. In Figure 10d it is shown that for $\theta_Y$ = 135° the droplet stays partially suspended even for the thinnest dielectric layer case that we have considered.

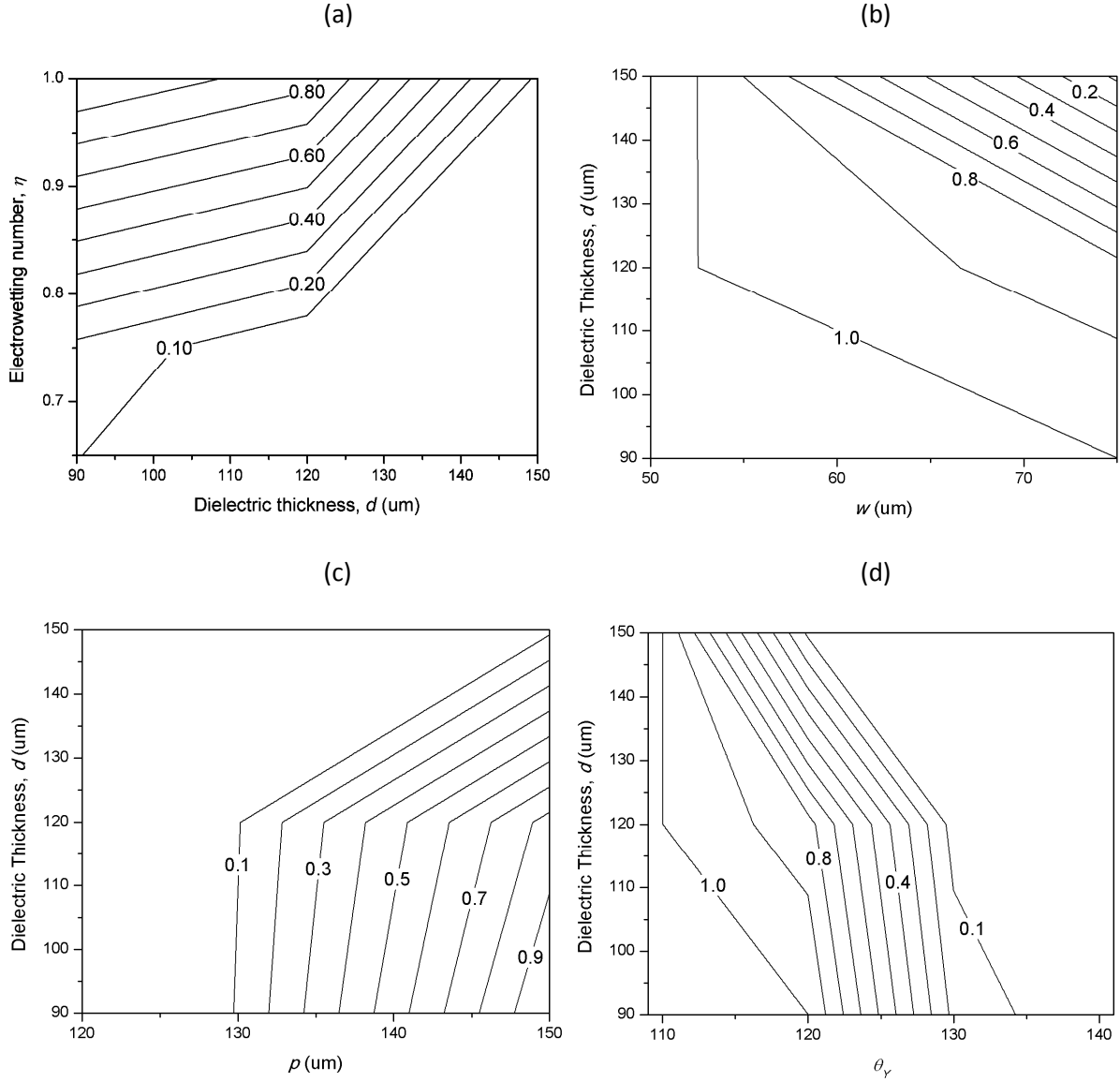

**Figure 10:** Contour plot presenting the fraction of the grooves filled with liquid (number of filled grooves normalized by the total number of grooves covered by the droplet) as a function of the dielectric thickness (ranging from $d$ = 90 um to $d$ = 150 um) and (a) the electrowetting number (ranging from $\eta$ = 0.5 to $\eta$ = 1) for $w$ = 75 um, $p$ = 150 um and θ$_Y$ = 120°, (b) the width of the stripes (ranging from $w$ = 90 um to $w$ = 150 um) for $\eta$ = 1, $p$ = 150 um and θ$_Y$ = 120°, (c) the period of the solid structures (ranging from $p$ = 105 um to $p$ = 150 um) for $\eta$ = 1, $w$ = 75 um and θ$_Y$ = 120°, (d) the Young contact angle (ranging from θ$_Y$ = 110° to θ$_Y$ = 140°) for $\eta$ = 1, $w$ = 75 um and $p$ = 150 um. The isoline where the fraction equals to 1 indicates a Wenzel state (where all the grooves covered by the droplet have been filled) whereas a lower fraction value represents a Cassie-like wetting state. The remaining geometric parameters of the structured solid surface are: $h$ = 75 um and $r$ = 15 um.

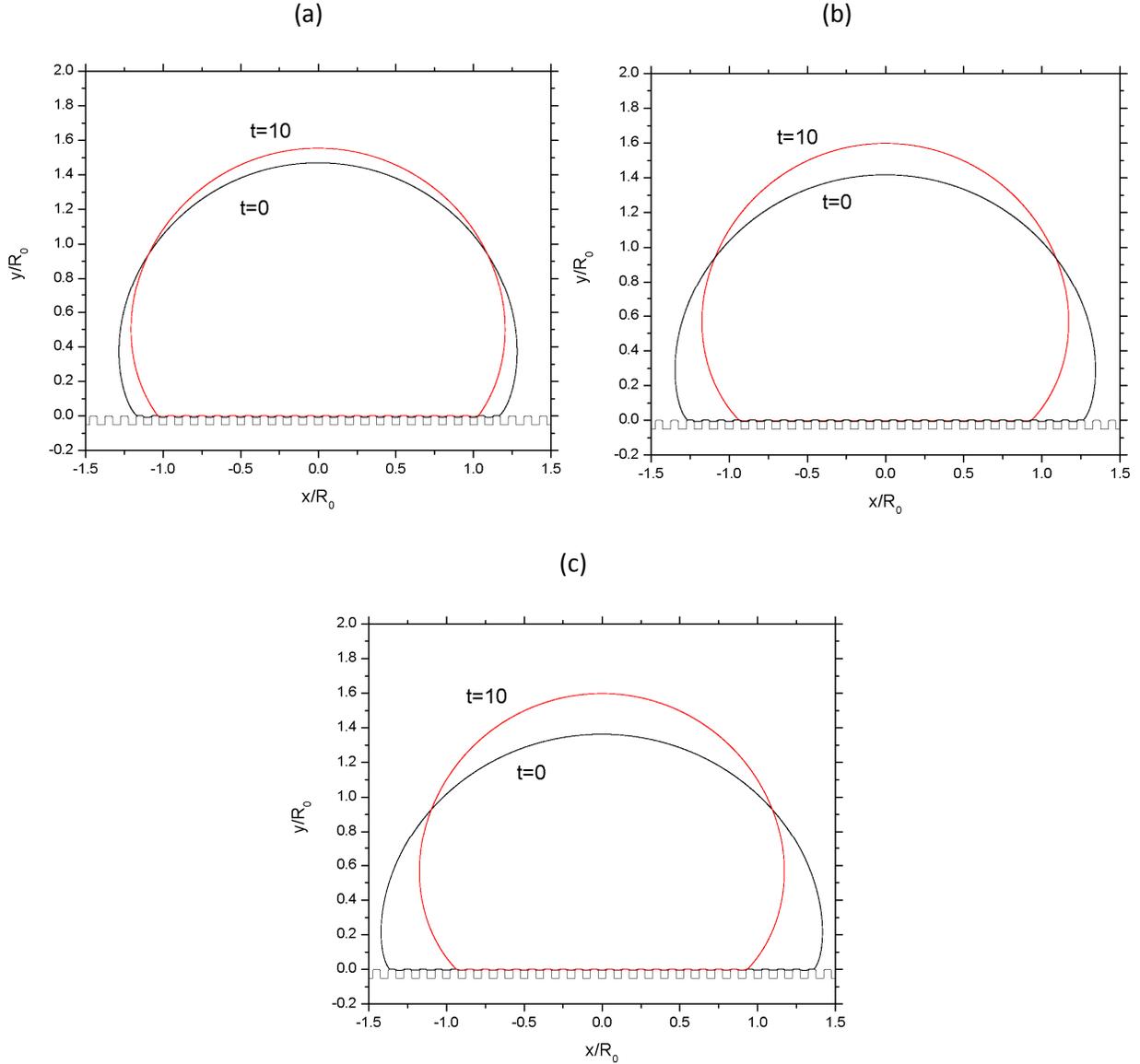

**Figure 11:** Initial and final state of the droplet when the electric field is turned off. The geometric parameters of the structured solid surface are: $h$ = 75 um, $w$ = 75 um, $p$ = 150 um and $r$ = 15 um and $\theta_Y$ = 140° and the dielectric layer has thickness (a) $d$ = 90 um, (b) $d$ = 150 um, (c) $d$ = 300 um At $t/t_c$ = 0 the droplet is at equilibrium with the electrowetting number equal to $\eta$ = 1 and at $t/t_c$ = 0⁺ the electrowetting number is set to $\eta$ = 0. At the $t/t_c$ = 10 the droplet has reached its new equilibrium state (video clips of the apparent contact angle reversibility, for all the three cases, are included in the supplementary material).

*Feasibility of reversible electrowetting – effect of contact angle hysteresis*

So far, we have examined the effect of the various characteristics of a structured solid surface on the resulting wetting state of the droplet when it is subjected to a voltage. Our parametric study helped us to identify under which conditions the droplet remains in Cassie-Baxter state and thus maintains an increased mobility which is an important condition in order to achieve fully reversible electrowetting. However, a question that arises is whether the droplet retracts to its initial shape when the voltage is turned off. In other words, whether the contact angle modification is reversible. To answer this question, we have selected the case of the substrate that can accommodate all the possible wetting states (Cassie-Baxter, Wenzel and mixed) where $h$ = 75 um, $w$ = 75 um, $p$ = 150 um and $r$ = 15 um (as shown in Figure 2a) and $\theta_Y$ = 140°; such an effective Young contact angle can be commonly observed on superhydrophobic surfaces due to the dual-scale topography [37] (the apparent contact angle in this case is greater than 160°). In Figure 11, we examine three different cases which correspond to dielectric layers of varying thickness ($d = 90, 150, 300$ um). At $t$ = 0 the electrowetting number is $\eta = 1$ and the droplet is at equilibrium, while at $t$ = 0⁺ the electric field is turned off ($\eta = 0$). Initially we observe that the droplet actually retracts back, close to its initial shape, when the applied voltage is removed (video clips of the apparent contact angle reversibility are included in the supplementary material). As we have already seen in Figure 7, the extend of spreading of the droplet when a voltage is applied increases with the thickness of the dielectric layer. Thus, there is significant difference in the initial stretching between the thinner (Figure 11a), and the thickest dielectric layer (Figure 11c). When the voltage is turned off, the initial stretching due to the formerly applied voltage cannot be supported and the capillary force acts now as a spring resulting in the retraction of the droplet. In order for the droplet to be able to retract, the driving force should also be able to overcome the contact angle hysteresis. This is reflected in the resulting state of equilibrium in these three cases where we observe that as the initial stretching of the droplet increases (with increasing dielectric thickness), the apparent contact angle of the retracted state increases because of the higher amount of momentum that the droplet gains due to the action of the capillary force. These results are in direct agreement with the experimental observations by [1, 18].

## Conclusions

In this work we investigated the impact of the solid topography and material wettability on the electrostatically-induced reversible wetting, at superhydrophobic surfaces. In our recent work [1, 18], we have showed that reversible wetting modification is only feasible when the dielectric thickness is sufficiently large, however, our argument has been only tested on a particular solid geometry. Here, we performed detailed electrowetting simulations on three different solid structures with varying stripes width ($w$ in Figure 2), distance ($p$ in Figure 2), material wettability, $\theta_Y$, as well as dielectric thickness, $d$. By employing a recently proposed computational scheme [26, 27], we have managed to predict collapse (Cassie-Baxter to Wenzel transitions) which makes the reversible wetting modification unachievable due to the enormous contact angle hysteresis. By avoiding collapse transitions and by increasing the Young contact angle (and thus reducing the contact angle hysteresis) we have demonstrated the retraction of the droplet, to its initial wetting state, when the applied voltage has been removed. Our main contribution here is that we are now able to define the critical dielectric thickness, for a particular solid geometry, beyond which no collapse transitions occur. Apart from the collapse transition occurrence we have also concluded that the contact angle hysteresis of the substrate is an equally important parameter for achieving reversible electrowetting. Such a finding is extremely important for designing modern miniaturized devices (e.g. lab-on-a-chip) where the liquid-solid adhesion can be dynamically controlled.

Undoubtedly, designing superhydrophobic surfaces with both low contact angle hysteresis and high mechanical robustness is a tedious task and it is a subject of ongoing research. Small scale asperities, like nanowires with high aspect ratio, for example, seem to be advantageous for impalement resistance and thus for electrowetting reversibility but their mechanical strength is poor. Future work focuses on performing more realistic three-dimensional simulations, for further investigating the geometric characteristics effect on wetting electrostatically-induced wetting reversibility and finally comparing our results with experimental measurements.

## Acknowledgments

The authors kindly acknowledge funding from the European Research Council under grant agreement no. 755412 (project HYDROPHO-CHEAP: COMMERCIALIZATION OF A NOVEL METHOD FOR FABRICATING CHEAP TAILOR-MADE SUPERHYDROPHOBIC SURFACES).